\documentclass{article}
\usepackage{spconf,amsmath,graphicx,hyperref}
\ninept

\usepackage{cite}
\usepackage{bm}
\usepackage{amsmath,amssymb,amsfonts}
\usepackage{algorithmic}
\usepackage{graphicx}
\usepackage{textcomp}
\usepackage{xcolor}
\usepackage{xspace}
\usepackage{url}
\usepackage{tabularx, booktabs,}
\usepackage{multicol, multirow}

\def\lcapi{$\mathcal{L}_\textrm{CA-PI-SDR}$\xspace}
\def\lca{$\mathcal{L}_\textrm{CA-SDR}$\xspace}
\def\lpi{$\mathcal{L}_\textrm{PI-SDR}$\xspace}

\def\mcapi{\texttt{CA-PI-SDRi}\xspace}

\def\dupset{\emph{DupSet}\xspace}
\def\nodupset{\emph{NoDupSet}\xspace}
\def\total{\emph{Total}\xspace}

\title{CLASS-AWARE PERMUTATION-INVARIANT SIGNAL-TO-DISTORTION RATIO FOR SEMANTIC SEGMENTATION OF SOUND SCENE WITH SAME-CLASS SOURCES}
\name{Binh Thien Nguyen,
Masahiro Yasuda,
Daiki Takeuchi, 
Daisuke Niizumi, 
and 
Noboru Harada\thanks{This work was partially supported by JST Strategic International Collaborative Research Program (SICORP), Grant Number JPMJSC2306, Japan.}}
\address{NTT, Inc.}
\begin{document}
%\ninept
%
\maketitle
\begin{abstract}
To advance immersive communication, the Detection and Classification of Acoustic Scenes and Events (DCASE) 2025 Challenge recently introduced Task 4 on Spatial Semantic Segmentation of Sound Scenes (S5).
An S5 system takes a multi-channel audio mixture as input and outputs single-channel dry sources along with their corresponding class labels.
Although the DCASE 2025 Challenge simplifies the task by constraining class labels in each mixture to be mutually exclusive, real-world mixtures frequently contain multiple sources from the same class.
The presence of duplicated labels can significantly degrade the performance of the label-queried source separation (LQSS) model, which is the key component of many existing S5 systems, and can also limit the validity of the official evaluation metric of DCASE 2025 Task 4.
To address these issues, we propose a class-aware permutation-invariant loss function that enables the LQSS model to handle queries involving duplicated labels.
In addition, we redesign the S5 evaluation metric to eliminate ambiguities caused by these same-class sources.
To evaluate the proposed method within the S5 system, we extend the label prediction model to support same-class labels.
Experimental results demonstrate the effectiveness of the proposed methods and the robustness of the new metric on mixtures both with and without same-class sources.

\end{abstract}


\vspace{-2mm}
\section{INTRODUCTION}
\label{sec:intro}
Immersive communication has become more practical and widely studied in recent years \cite{ivas, masa, immersive1}, particularly with the introduction of the Immersive Voice and Audio Services (IVAS) codec \cite{ivas} and the novel parametric spatial audio format, Metadata-Assisted Spatial Audio (MASA) \cite{masa}. % immersive2
Its core technologies include decomposing complex spatial sound scenes into individual sound sources along with metadata characterizing their properties.
To support progress in immersive communication, the Detection and Classification of
Acoustic Scenes and Events (DCASE) 2025 Challenge Task 4 (DCASE25T4) \cite{dcase2025}, titled ``Spatial Semantic Segmentation of Sound Scenes (S5)\footnote{https://dcase.community/challenge2025/\#task4}," focuses on detection and separation of sound sources.
An S5 system processes a multi-channel audio mixture and produces separated single-channel dry sources with their corresponding class labels.

The baseline system for DCASE25T4 \cite{baseline} uses a two-stage approach, comprising an audio tagging (AT) model that predicts source labels from the mixture, followed by a label-queried source separation (LQSS) model that extracts the corresponding sources.
Various systems \cite{dcase25_multi1, dcase25_multi2, dcase25_ite1, dcase25_ite2, dcase25_ite3, dcase25_ite_com, dcase25_com1} have been proposed, most of which utilize similar multi-stage strategies.
In addition to improving the model architectures, a common approach among these systems \cite{dcase25_ite1, dcase25_ite2, dcase25_ite3, dcase25_ite_com} is to iteratively alternate between label prediction and source separation, allowing the outputs of each stage to mutually refine one another.
An alternative strategy \cite{dcase25_ite_com, dcase25_com1} employs a single model to jointly perform label prediction and source separation, trained using permutation-invariant training (PIT) \cite{pit}. However, this approach underperforms compared to multi-stage strategies.

\begin{figure}[t]
\begin{center}
\includegraphics[width=\columnwidth]{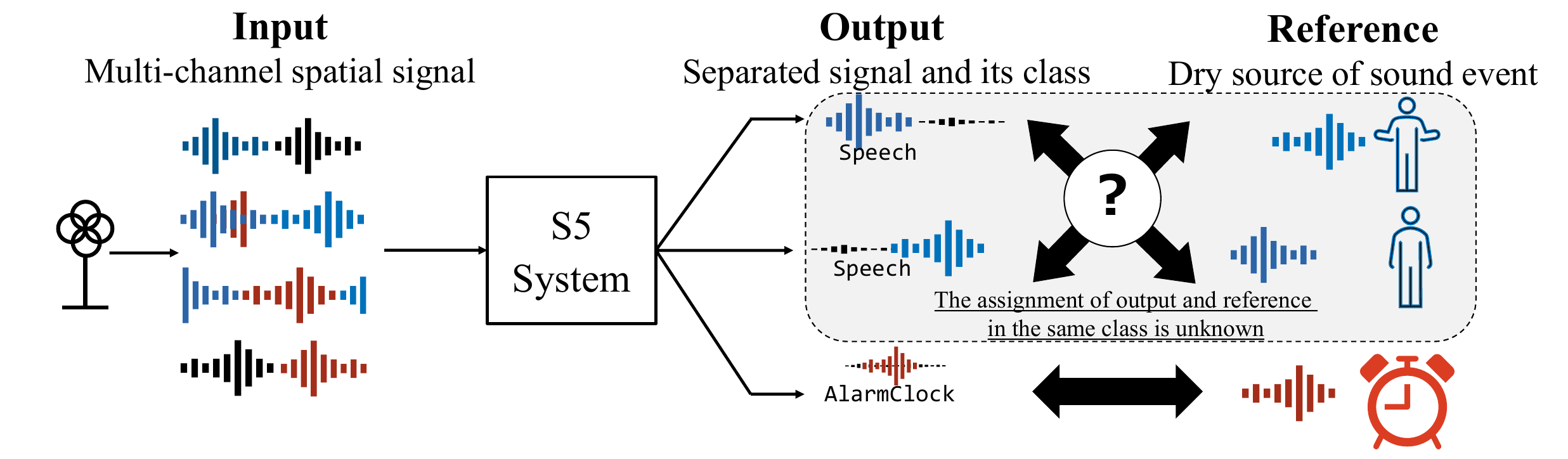}
\end{center}
\vspace{-0.7cm}
\caption{Overview of S5 task with same-class sources.}\label{fig:overview}
\vspace{-0.4cm}
\end{figure}

In the context of DCASE25T4, it is assumed that source labels in a mixture are mutually exclusive, meaning that each label can appear at most once per mixture.
However, in practical scenarios, it is common for mixtures to contain multiple sound sources of the same class, such as two people speaking simultaneously, as illustrated in Fig.~\ref{fig:overview}.
In such cases, correctly separating these sources can be challenging for most existing S5 systems, and may even be infeasible for systems that only process distinct labels.
Label duplication is a known problem in sound event localization and detection \cite{seld1, seld2}; however, they have not been thoroughly investigated in LQSS.
LQSS (also known as target sound extraction) systems \cite{tse1, tse2, tse4, resunet} typically process a single label per query, with multiple sources being extracted across separate queries. % tse3
With this approach, these systems may fail to correctly separate same-class sources, instead extracting either a single source or a mixture of all sources associated with that class.
The DCASE25T4 baseline presents another approach for LQSS, ResUNetK, which queries multiple sources simultaneously by concatenating multiple labels into a single query. However, since the order of the output sources is determined by the order of the labels, duplicated labels create ambiguity, negatively affecting the training process.
Blind source separation techniques, such as PIT, may mitigate this problem.
However, they do not leverage prior label information and have been shown to yield lower results than LQSS-based systems in DCASE25T4, as previously discussed.
In addition to impacting system performance, same-class sound sources in the mixture also invalidates the official DCASE25T4 metric for evaluating S5 systems, class-aware signal-to-distortion ratio improvement (CA-SDRi) \cite{dcase2025}, due to the confusion caused by duplicated labels.

In this paper, we address the challenges introduced by same-class sources in LQSS and S5.
Specifically, we introduce a loss function for training LQSS models that resolves ambiguities arising from duplicated label queries.
We demonstrate the proposed approach on the DCASE25T4 baseline systems, with modifications to the AT model.
These techniques are readily transferable to other LQSS and S5 systems.
In addition, we introduce a metric for jointly evaluating label prediction and source separation in S5 systems, applicable to mixtures with or without same-class sources.
The source code is released as part of the baseline system and evaluation metric for the DCASE 2026 Challenge Task.

% \vspace{-0.2cm}
\section{BASELINE S5 SYSTEM}
\label{sec:relwork}
% \vspace{-0.1cm}

\begin{figure}[t]
\begin{center}
\includegraphics[trim={0 0 0 0},clip,scale=0.96]{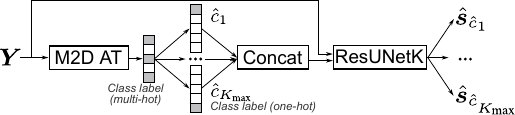}
\end{center}
\vspace{-0.4cm}
\caption{Baseline ResUNetK-based S5 Systems.}\label{fig:baseline}
\vspace{-0.4cm}
\end{figure}

This section describes a baseline S5 system of DCASE25T4 \cite{baseline}, the diagram of which is shown in Fig.~\ref{fig:baseline}.
The input to the S5 system is $\bm{Y} \in \mathbb{R}^{M \times T}$, an $M$-channel time-domain mixture signal of length $T$.
The output consists of $K$ labels, $C = (c_1, \dots, c_K)$, together with their corresponding separated single-channel waveforms at a reference microphone, $S = (\bm{s}_{c_1},\dots,\bm{s}_{c_K})$, where $\bm{s}_{c_k} \in \mathbb{R}^{T}$, and $K$ may vary from $1$ to $K_\textrm{max}$.
We denote the estimated labels and waveforms by $\hat{C} = (\hat{c}_1, \ldots, \hat{c}_{\hat{K}})$ and $\hat{S} = (\hat{\bm{s}}_{\hat{c}_1}, \ldots, \hat{\bm{s}}_{\hat{c}_{\hat{K}}})$, respectively. The system comprises two stages, detailed below.

The first stage predicts the labels $\hat{C}$ using the Masked Modeling Duo (M2D) AT model \cite{m2d}, which employs an M2D backbone—a self-supervised model pre-trained on AudioSet—to extract high-quality audio representations.
The M2D AT model takes the reference channel of the mixture $\bm{Y}$ as input and outputs a multi-hot vector. This multi-hot vector implicitly determines the estimated number of sources in the mixture, $\hat{K}$. It is then split into $K_\textrm{max}$ one-hot vectors (with zero padding if $\hat{K} < K_\textrm{max}$), concatenated, and forwarded to the second stage.

In the second stage, ResUNetK is used to extract the sources corresponding to the labels generated in the first stage.
ResUNetK is an extension of ResUNet \cite{resunet}, an LQSS model.
While ResUNet extracts one source at a time using a single-label query, ResUNetK processes a multi-label query and extracts multiple sources simultaneously.
The output sources are aligned with the reference sources based on the order of the input labels, and the SDR loss is computed on these aligned pairs to optimize the model.

Evaluating S5 systems requires jointly assessing label prediction and source separation.
\cite{baseline} introduced the CA-SDRi metric, which has been adopted as the official evaluation metric for DCASE25T4.
CA-SDRi evaluates system performance by first aligning estimated sources with their corresponding references based on predicted labels. The SDRi is then computed only for sources with correctly predicted labels, while penalty values are assigned to incorrect predictions.
This ensures that the metric jointly captures both source separation quality and label prediction accuracy.

While the baseline system achieves comparable performance on the S5 task for mixtures with distinct labels, it fails to separate sources of the same class, due to the inability of the multi-hot vector output in the first stage to represent duplicated labels.
Even when duplicated labels can be predicted correctly, they still disrupt the label-based estimated-to-reference source alignment process, which impairs both the loss function used to train the LQSS and the CA-SDRi metric for evaluating S5 systems.
Possible solutions to these issues are presented in the following section.

% \vspace{-0.2cm}
\section{PROPOSED S5 SYSTEM}
\label{sec:proposed}
% \vspace{-0.2cm}
\subsection{Audio tagging model}
\begin{figure}[t]
\begin{center}
\includegraphics[trim={0 0 0 0},clip,scale=1]{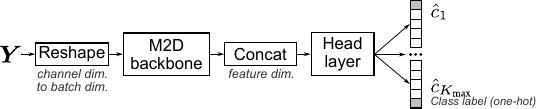}
\end{center}
\vspace{-0.6cm}
\caption{Modified M2D AT system.}\label{fig:m2d_mod}
\vspace{-0.2cm}
\end{figure}

Figure~\ref{fig:m2d_mod} illustrates the modified M2D AT architecture.
Instead of predicting a multi-hot vector, we adopt the track-based approach as in \cite{seld1}, which outputs multiple one-hot vectors, thereby enabling repeated label prediction.
The model is trained using PIT with cross-entropy loss.
We further exploit spatial information by processing all waveform channels of $\bm{Y}$ instead of relying on a single-channel input as in \cite{baseline}.
Specifically, we reshape the channel dimension into the batch dimension, enabling the M2D backbone to treat the multi-channel signal as multiple single-channel signals, extract features independently, and then concatenate them along the feature dimension before passing them to the head layer.

\vspace{-0.3cm}
\subsection{Source separation model} \label{ssec:prop_ss}
For the separation model architecture, we adopt the ResUNetK in \cite{baseline} and modify the loss function.
The order of input labels is also used to align the estimated and reference sources to calculate the SDR loss function.
To handle cases where same-class sources exist (i.e., duplicated input labels), we apply PIT to map these same-class output sources to the corresponding reference sources in a manner that minimizes the average loss.
This method can also be interpreted as standard PIT, except that permutations are permitted only among same-class sources.
The loss function is formulated as follows.

Let $\mathfrak{S}_K$ denote a set of all permutations of $K$ indices $\{1, \dots, K\}$,
\begin{equation}
\mathfrak{S}_K = \bigl\{ (i_1, \dots, i_K) \;\big|\; i_j \in \{1, \dots, K\},\; i_j \neq i_m \text{ for all } j \neq m \bigr\}.
\end{equation}
For same-class permutation constraints, given a label sequence $C$, we consider only a set of permutations that reorder elements within each class, denoted $\mathfrak{S}^C_K$. That is, after permutation, the class labels of elements in the sequence remain unchanged.
$\mathfrak{S}^C_K$ is defined as
\begin{equation}
	\mathfrak{S}^C_K = \{\pi \in \mathfrak{S}_K \mid c_{\pi(k)} = c_k \;\; \forall k \in \{1,\dots,K\}\}.
\end{equation}

The loss function for training the separation model, namely class-aware permutation-invariant SDR (CA-PI-SDR), is defined by minimizing the reconstruction error over permutations in $\mathfrak{S}^C_K$ as
\begin{equation}\label{eq:loss}
	\mathcal{L}_\textrm{CA-PI-SDR} = \min_{\pi \in \mathfrak{S}^C_K} \left(-\frac{1}{K} \sum_{k=1}^K \textrm{SDR}(\hat{\bm{s}}_{c_{\pi(k)}}, \bm{s}_{c_k}) \right),
\end{equation}
\begin{equation}\label{sq:sdr}
	\textrm{SDR}(\hat{\bm{s}}, \bm{s}) 
	= 10\log_{10} \left( \frac{\|\bm{s}\|^2}{\|\bm{s} - \hat{\bm{s}}\|^2} \right).
\end{equation}
As with \cite{baseline}, the oracle labels are provided during training.

\subsection{Evaluation metric}
The proposed metric follows a similar principle as the loss function in Section \ref{ssec:prop_ss}, using label-based source aligning and a permutation-invariant objective to handle label duplicates.
However, unlike during training, which relies on oracle labels, the number and labels of estimated and reference sources during evaluation may differ depending on label predictions from the first stage.
Additional processing is therefore required to handle these cases, as described below.

\subsubsection{Notations}
We first introduce some notations used in the metric.
Let $\mathcal{C} = \textrm{set}(C)$ and $\hat{\mathcal{C}} = \textrm{set}(\hat{C})$ be the sets of unique labels in $C$ and $\hat{C}$, respectively.
A label from the union of reference and predicted sets is denoted by $\bar{c}\in \mathcal{C} \cup \hat{\mathcal{C}}$.
The collections of reference and estimated waveforms associated with label $\bar{c}$ are defined as
\vspace{-0.1cm}
\begin{equation}
\vspace{-0.1cm}
\begin{aligned}
	S^{\bar{c}} = (\bm{s}_{c_k} \in S \mid c_k = \bar{c}) = (\bm{s}^{\bar{c}}_1,\dots, \bm{s}^{\bar{c}}_{|S^{\bar{c}}|}),\\
	\hat{S}^{\bar{c}} = (\hat{\bm{s}}_{\hat{c}_k} \in \hat{S} \mid \hat{c}_k = \bar{c}) = (\hat{\bm{s}}^{\bar{c}}_1,\dots, \hat{\bm{s}}^{\bar{c}}_{|\hat{S}^{\bar{c}}|}),
\end{aligned}
\end{equation}
where $|\cdot|$ denotes the number of elements in the collection, and $\bm{s}^{\bar{c}}_i$ and $\hat{\bm{s}}^{\bar{c}}_i$ represent the $i$-th elements of $S^{\bar{c}}$ and $\hat{S}^{\bar{c}}$, respectively.
The number of true positive (TP), false negative (FN), and false positive (FP) predictions  for label $\bar{c}$ can be computed as
\vspace{-0.1cm}
\begin{equation}
\vspace{-0.1cm}
\begin{aligned}
N_{\mathrm{TP}}^{\bar{c}} = \min\bigl(|S^{\bar{c}}|, &|\hat{S}^{\bar{c}}|\bigr), \quad
N_{\mathrm{FN}}^{\bar{c}} = \bigl(|S^{\bar{c}}| - |\hat{S}^{\bar{c}}|\bigr)_+,\\
N_{\mathrm{FP}}^{\bar{c}} &= \bigl(|\hat{S}^{\bar{c}}| - |S^{\bar{c}}|\bigr)_+,
\end{aligned}
\end{equation}
with $(x)_+ = \max(0,x)$, and the total number of true and false predictions is calculated as
\vspace{-0.1cm}
\begin{equation}
\vspace{-0.1cm}
N^{\bar{c}} = N_{\mathrm{TP}}^{\bar{c}} + N_{\mathrm{FN}}^{\bar{c}} + N_{\mathrm{FP}}^{\bar{c}} 
    = \max\bigl(|S^{\bar{c}}|, |\hat{S}^{\bar{c}}|\bigr).
\end{equation}
It is worth noting that for each $\bar{c}$, either $N_{\mathrm{FN}}^{\bar{c}}$, $N_{\mathrm{FP}}^{\bar{c}}$, or both are zero.

Let $\mathfrak{S}_{K,L}$ denote the set of all permutations of $L$ distinct indices chosen from $\{1, \dots, K\}$, defined as
\begin{equation}
\mathfrak{S}_{K,L} = \bigl\{ (i_1, \dots, i_L) \;\big|\; i_j \in \{1, \dots, K\},\; i_j \neq i_m \text{ for all } j \neq m \bigr\},
\end{equation}
Similarly, $\mathfrak{C}_{K,L}$ denotes the set of all combinations (i.e., without permutation) of $L$ distinct indices chosen from $\{1, \dots, K\}$ as
\vspace{-0.1cm}
\begin{equation}
\vspace{-0.1cm}
\mathfrak{C}_{K,L} = \Bigl\{ (i_1, \dots, i_L) \;\Big|\; 1 \le i_1 < i_2 < \dots < i_L \le K \Bigr\}.
\end{equation}

\subsubsection{Class-aware permutation-invariant metric}
The metric is first calculated for each label $\bar{c}$. Given $S^{\bar{c}}$ and  $\hat{S}^{\bar{c}}$, $N_{\mathrm{TP}}^{\bar{c}}$ sources selected from $S^{\bar{c}}$ and $N_{\mathrm{TP}}^{\bar{c}}$ from $\hat{S}^{\bar{c}}$ are paired using a permutation-invariant objective to maximize the average metric.
For selecting and aligning these sources, we consider all the permutations of the estimated sources using $\mathfrak{S}_{|\hat{S}^{\bar{c}}|, N^{\bar{c}}_\textrm{TP}}$, while it suffices to select the reference sources using $\mathfrak{C}_{|S^{\bar{c}}|, N^{\bar{c}}_\textrm{TP}}$ without any permutation.
The remaining $N_{\mathrm{FN}}^{\bar{c}}$ sources in $S^{\bar{c}}$ and $N_{\mathrm{FP}}^{\bar{c}}$ in $\hat{S}^{\bar{c}}$ are treated as false predictions and are penalized.
The metric component corresponding to $\bar{c}$ is defined as
\vspace{-0.3cm}
\begin{equation}\label{eq:metric_comp}
\vspace{-0.1cm}
\begin{split}
P^{\bar{c}}
	=
		N^{\bar{c}}_\textrm{FN}\mathcal{P}_\textrm{FN} +
		N^{\bar{c}}_\textrm{FP}\mathcal{P}_\textrm{FP} + 
		\max_{\substack{
			\sigma \in \mathfrak{S}_{|\hat{S}^{\bar{c}}|, N^{\bar{c}}_\textrm{TP}}\\
			\pi \in \mathfrak{C}_{|S^{\bar{c}}|, N^{\bar{c}}_\textrm{TP}}
			}} 
		\sum_{i = 1}^{N^{\bar{c}}_\textrm{TP}} \textrm{SDRi}(\hat{\bm{s}}_{\sigma (i)}^{\bar{c}}, \bm{s}_{\pi (i)}^{\bar{c}}, \bm{y}) ,
\end{split}
\end{equation}
where $\mathcal{P}_\textrm{FN}$ and $\mathcal{P}_\textrm{FP}$ are penalty values for FN and FP, respectively, both of which are set to $0$ following \cite{baseline}.
$\bm{y}$ is the waveform at the reference channel of $\bm{Y}$, used for calculating the SDRi as
\vspace{-0.1cm}
\begin{equation}
	\textrm{SDRi}(\hat{\bm{s}}, \bm{s}, \bm{y})
	= \textrm{SDR}(\hat{\bm{s}}, \bm{s}) - \textrm{SDR}(\bm{y} , \bm{s}).
\end{equation}

The class-aware permutation-invariant SDRi (CA-PI-SDRi) metric is the average of $P^{\bar{c}}$ over all the true and false predictions as
\begin{equation} \label{eq:metric}
	\textrm{CA-PI-SDRi}(\hat{S}, S, \hat{C}, C, \bm{y}) =
	\frac{1}{\sum_{\bar{c} \in \mathcal{C} \cup \hat{\mathcal{C}}} N^{\bar{c}}} 
	\sum_{\bar{c} \in \mathcal{C} \cup \hat{\mathcal{C}}} P^{\bar{c}}.
\end{equation}
We remark that CA-PI-SDRi reduces to CA-SDRi in \cite{baseline} when all source labels in the mixture are mutually exclusive, while also supporting mixtures containing multiple sources of the same class.

It is worth noting that, while  $\mathcal{P}_\textrm{FN}=\mathcal{P}_\textrm{FP}=0$ contributes nothing to $\mathcal{P}^{\bar{c}}$ in (\ref{eq:metric_comp}), the final metric value in (\ref{eq:metric}) is still penalized due to the increase in $N^{\bar{c}}$ caused by false predictions.
In addition, $\mathcal{P}_\textrm{FN}$ and $\mathcal{P}_\textrm{FP}$ can be chosen to decrease or increase the penalty, or alternatively scaled based on the falsely estimated sources.

\begin{table}[t!]
\caption{Performance of M2D AT Models}
\label{tab:at}
\centering
\begin{tabular}{
>{\centering\arraybackslash}p{0.6cm}
>{\centering\arraybackslash}p{0.7cm}
>{\centering\arraybackslash}p{1.1cm}
>{\centering\arraybackslash}p{0.7cm}
>{\centering\arraybackslash}p{0.7cm}
>{\centering\arraybackslash}p{1.1cm}
>{\centering\arraybackslash}p{0.7cm}
}
\toprule
\multirow{2}{*}{\parbox{0.6cm}{\centering Input\\Chan.}} 
& \multicolumn{3}{c}{Source accuracy [\%]} 
& \multicolumn{3}{c}{Mixture accuracy [\%]} \\
\cmidrule(lr){2-4} \cmidrule(lr){5-7}
& {\dupset} & {\nodupset} & {\total}
& {\dupset} & {\nodupset} & {\total} \\
\midrule
4
& \textbf{77.9} & \textbf{79.4} & \textbf{78.7}
& \textbf{55.4} & \textbf{68.3} & \textbf{63.2} \\
1
& 74.6 & 77.7 & 76.3
& 48.6 & 66.9 & 59.6 \\
\bottomrule
\end{tabular}
\end{table}

\begin{figure}[t]
\begin{center}
\includegraphics[trim={0 0 0 0},clip,scale=1]{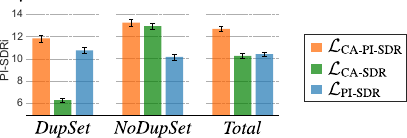}
\end{center}
\vspace{-0.5cm}
\caption{Performance ResUNetK trained with various loss functions.}\label{fig:results_sep}
\vspace{-0.3cm}
\end{figure}

\begin{figure*}[t]
\begin{center}
\includegraphics[trim={0 0 0 0},clip,scale=0.9]{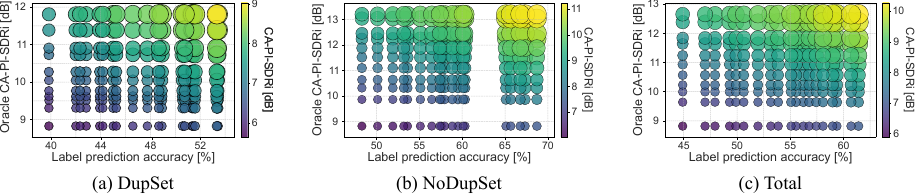}
\end{center}
\vspace{-0.5cm}
\caption{Performance of various S5 systems. The x-axis shows the accuracy of AT models in the first stage, while the y-axis shows the performance of the separation models in the second stage when provided with oracle labels. The size and color of each point indicate the overall CA-PI-SDRi achieved by the full system.}\label{fig:results_capi}
\vspace{-0.3cm}
\end{figure*}

\section{EXPERIMENTS AND RESULTS}
\label{sec:experiments}
\subsection{Experimental setting}
The multi-channel mixtures were synthesized by convolving target and interfering sound sources with first-order Ambisonics room impulse responses (RIRs) and summing them together with multichannel background noise. This followed the procedure described in \cite{dcase2025}, which employed a modified version of the SpatialScaper toolkit \cite{spatialscaper}.

All sources had a duration of $T = 10$ seconds and were sampled at 32 kHz.
Each mixture contained between one and $K_\textrm{max} = 3$ target sources, with individual sources having signal-to-noise ratios (SNRs) ranging from 5 to 20 dB.
Additionally, each mixture included between 0 and 2 interference sources, with SNRs ranging from 0 to 15 dB.
The target dry sources $\bm{s}_{c_k}$ were obtained by convolving the original sources with the direct-path component of the impulse response ($-6$ to $50$\,ms around its first peak).
For target sources belonging to the same class, we ensured that their directions of arrival differed by at least 60 degrees. The reference channel was set to 1, corresponding to the omnidirectional channel.

For the target and interference sources, we adopted the train, validation, and test splits defined in \cite{semhear}.
For the RIRs and background noise, we used the data from \cite{dcase2025}, which includes samples from FOA-MEIR \cite{foameir} dataset as well as newly collected data for DCASE25T4.

We generated a test dataset of 3,000 mixtures, divided into two subsets. \dupset comprises 1,200 mixtures containing same-class target sources (600 with 2 sources, 600 with 3), while \nodupset comprises 1,800 mixtures in which all target sources belong to distinct classes (600 with 1 source, 600 with 2, 600 with 3).
Validation mixtures were generated from the validation split following a similar source distribution.
Training mixtures were synthesized dynamically during training, where the models were trained using mixtures both with and without same-class sound sources.

Aside from the modifications described in Section \ref{sec:proposed}, the remaining components of the S5 systems in the experiments were similar to those in \cite{baseline}.
All models were trained on 4 RTX 3090 GPUs using the Adam optimizer. The separation models were trained with a batch size of 4 for 500 epochs, using a learning rate of $10^{-4}$. Training of the AT model proceeded in two steps: first, the head was trained for 300 epochs with a batch size of 16 and a learning rate of $10^{-3}$; second, the head and two M2D blocks were fine-tuned for 200 epochs with a batch size of 8 and a learning rate of $10^{-5}$.

\vspace{-0.1cm}
\subsection{Audio tagging results}
Table \ref{tab:at} presents the accuracies of the proposed M2D AT models.
Source accuracy is computed across all sources in all mixtures as the total number of TP divided by the total number of TP, FP and FN.
Mixture accuracy is counted per mixture: a mixture is considered correct only if all labels, including duplicated labels, are predicted correctly.
% Table \ref{tab:at} presents the accuracy of the proposed M2D AT model on \dupset, \nodupset, and the full test dataset (\total). For comparison, we also report the results of a single-channel model.
For both AT models with 1 and 4 input channels, performance on \nodupset exceeds that on the more challenging \dupset.
We also observe that using multi-channel input improves accuracy compared to single-channel input, particularly on \dupset.

\subsection{Assessment of loss functions for separation model}
In this experiment, we compare three loss functions for the source separation model in the second stage by training ResUNetK with each to evaluate their performance.
All three are calculated using SDR and differ only in how the estimated sources are aligned to the reference sources when computing the loss function.
These are
\begin{enumerate}
    \item \lcapi, the proposed loss function (\ref{eq:loss}), aligning sources using both input labels and permutation invariant objectives.
    \item \lca, the baseline class-aware loss function (as described in \cite{baseline}), aligning the sources based on the input labels. In the case of same-class source elements, it performs a random mapping. This loss function is equivalent to (\ref{eq:loss}), except that the permutation $\pi$ is selected randomly from $\mathfrak{S}^C_K$ rather than being selected to minimize the loss function.
    \item \lpi, the PIT-based loss function, typically used in blind source separation. This loss function is equivalent to (\ref{eq:loss}) with $\mathfrak{S}^C_K$ replaced by $\mathfrak{S}_K$, aligning the sources to minimize the loss regardless of the labels.
\end{enumerate}
We used PI-SDRi, an SDRi metric computed with a permutation-invariant objective, to compare the outputs of these models.

Results are shown in Fig.~\ref{fig:results_sep}.
We can see that while \lca performs well on \nodupset, its results drop significantly on \dupset. This is likely due to the random mapping of same-class sources, which confuses the model during training. We also observed that, when testing the model trained using \lca with queries containing duplicated labels, it often produces similar sources for the repeated labels rather than properly separating them.
In contrast, \lpi can mitigate the issue of duplicated class labels, illustrated by the comparable results on \nodupset. However, it does not exploit the available label information, resulting in notably lower performance than \lca on \nodupset.
By leveraging the advantages of both approaches, the proposed \lcapi not only maintains high performance on \nodupset, but also yields reasonable results on \dupset, resulting in the highest average performance.

\subsection{CA-PI-SDRi metric on various S5 systems}

This experiment assesses the efficacy of the proposed metric in jointly evaluating label prediction and source separation performance in S5.
Two stages of the proposed S5 model were trained with checkpoints saved at various epochs, resulting in multiple system variants with varying label accuracy and separation quality.

Figure~\ref{fig:results_capi} shows the performance of the first-stage AT model, the second-stage source separation model, and the complete S5 system combining both stages.
Results on both subsets of the test dataset exhibit a consistent pattern: \mcapi metric increases with higher label prediction accuracy along the x-axis and with higher source separation performance along the y-axis.
Accordingly, the best-performing S5 system is located in the top-right corner, reflecting both high label prediction accuracy and separation performance, while the worst-performing system appears in the bottom-left corner.
These results demonstrate that the proposed metric effectively reflects the overall performance of S5 systems, simultaneously capturing both label prediction and separation quality.

\section{Conclusions} \label{sec:conclusions}
\vspace{-0.1cm}
In this paper, we present a CA-PI-SDR loss function for training the LQSS model in the S5 system, with the AT model also modified to handle mixtures containing multiple sources of the same class.
We also propose an evaluation metric to address the confusion caused by duplicated labels.
The experimental results demonstrate the effectiveness of the proposed S5 system and support the relevance of the metric for mixtures both with and without same-class sound sources.

While the paper focuses on the separation model, the performance of the AT model remains limited due to the requirement of simultaneously estimating the number of sources and their labels, a task that becomes especially challenging when multiple sources belong to the same class.
In addition, although the modified AT model processes all the channels of the mixture, M2D backbone is still pre-trained on single channel input, which may limit the ability to extract spatial information.
Future work will explore methods to fully exploit multi-channel input and adapt pretrained models for this setting.
In addition, findings from DCASE25T4 suggest that the outputs of source separation and label prediction can refine each other through iterative schemes.
However, this strategy typically increases both the model size and processing time.
A promising direction for future work is to develop a unified model that jointly performs label prediction and sound source separation, while efficiently leveraging the complementary information between the two tasks.

% -------------------------------------------------------------------------
\bibliographystyle{IEEEbib}
\bibliography{refs}

\end{document}